%% file: main.tex
\documentclass[letterpaper,twocolumn,10pt]{article}
\usepackage{usenix-2020-09}

\usepackage{booktabs}
\usepackage{tikz}
\usepackage{comment}
\usepackage{amsmath}
\usepackage{caption}
\usepackage{subcaption}
\usepackage{graphicx}
\usepackage{float}
\usepackage{algorithm}
\usepackage{algorithmicx}
\usepackage[noend]{algpseudocode}
\algnewcommand{\LineComment}[1]{\State \(//\) #1}

\usepackage[normalem]{ulem}
\usepackage{tikz}
\usepackage{amsmath}
\usepackage{bbding}
\usepackage{todonotes}
\usepackage{filecontents}
\usepackage{enumitem}
\usepackage[bottom]{footmisc}
\usepackage{hyperref}
\usepackage{cleveref}
\definecolor{LightGray}{gray}{0.9}
\usepackage{pifont}
\usepackage{multirow}

\usepackage{xcolor}
\definecolor{FadedBanana}{RGB}{255,255,191}
\definecolor{DeepChalk}{RGB}{255,191,191}
\definecolor{FadedFlora}{RGB}{191,255,191}
\definecolor{DeepSnow}{RGB}{191,255,255}
\definecolor{SoapStone}{RGB}{218,218,218}
\definecolor{LightCayenneSixty}{RGB}{239,206,211}
\definecolor{LightCayenne}{RGB}{208,143,145}

\newcommand{\sysname}{{SAGE}}
\newcommand{\sota}{{FixedGSL}}

\makeatletter
\renewcommand{\paragraph}{%
  \@startsection{paragraph}{4}%
  {\z@}{0.25ex \@plus 1ex \@minus .2ex}{-1em}%
  {\normalfont\normalsize\bfseries}%
}

\usepackage{etoolbox}
\makeatletter
\patchcmd{\maketitle}
	{\@maketitle}
	{\@maketitle\vspace{-2em}}
	{}
	{}
\makeatother

\begin{document}
\date{}


\title{\Large \bf Towards Fast Setup and High Throughput of GPU Serverless Computing}

\author{
{\rm Han Zhao$^*$, Weihao Cui$^*$, Quan Chen$^{*}$, Shulai Zhang$^*$}\\
{\rm Zijun Li$^*$, Jingwen Leng$^*$, Chao Li$^*$, Deze Zeng$^{\diamond}$, Minyi Guo$^{*}$ }\\
$^*$Shanghai Jiao Tong University, $^\diamond$China University of Geosciences
}

\maketitle
\begin{abstract}

Integrating GPUs into serverless computing platforms is crucial for improving efficiency. However, existing solutions for GPU-enabled serverless computing platforms face two significant problems due to coarse-grained GPU management: long setup time and low function throughput.

To address these issues, we propose \sysname{}, a GPU serverless framework with fast setup and high throughput. First, based on the data \textit{knowability} of GPU function ahead of actual execution, \sysname{} first devises the parallelized function setup mechanism, which parallelizes the data preparation and context creation. In this way, \sysname{} achieves fast setup of GPU function invocations.Second, \sysname{} further proposes the sharing-based memory management mechanism, which shares the read-only memory and context memory across multiple invocations of the same function. The memory sharing mechanism avoids repeated data preparation and then unnecessary data-loading contention. As a consequence, the function throughput could be improved. Our experimental results show that \sysname{} reduces function duration by $11.3\times$ and improves function density by $1.22\times$ compared to the state-of-the-art serverless platform.
\end{abstract}


\input{contents/introduction}
\input{contents/related.tex}

\input{contents/motivation.tex}

\input{contents/design.tex}
\input{contents/manager.tex}
\input{contents/allocator.tex}
\input{contents/evaluation.tex}
\input{contents/conclusion.tex}

\bibliographystyle{unsrt}
\bibliography{bib}

\end{document}

%% file: contents/introduction.tex
\section{Introduction}

Emerging clouds start to employ GPUs for the cloud applications (e.g., personal recommendation~\cite{ke2022hercules}, medical service~\cite{parboil}). These applications often integrate AI or scientific computing components, which rely on GPUs for computation. Meanwhile, these cloud applications often experience unstable loads, which may have only a few to dozens of user requests per minute~\cite{serverless_ai, spillner2018faaster,eismann2020serverless,van2017spec}. 
In this case, if a traditional GPU-based inference system~\cite{wang2019distributed,carreira2018case,ishakian2018serving,ali2020batch,ali2022optimizing,li2022tetris} is adopted, the resources will be reserved for a long time, even if the load is low.
Serverless computing~\cite{azurefaasgpu,li2022help,eismann2020serverless} has been proven efficient for handling relatively low and unstable loads, benefitting from the high scalability.

\begin{figure}
	\centering
	\includegraphics[width=0.9\linewidth]{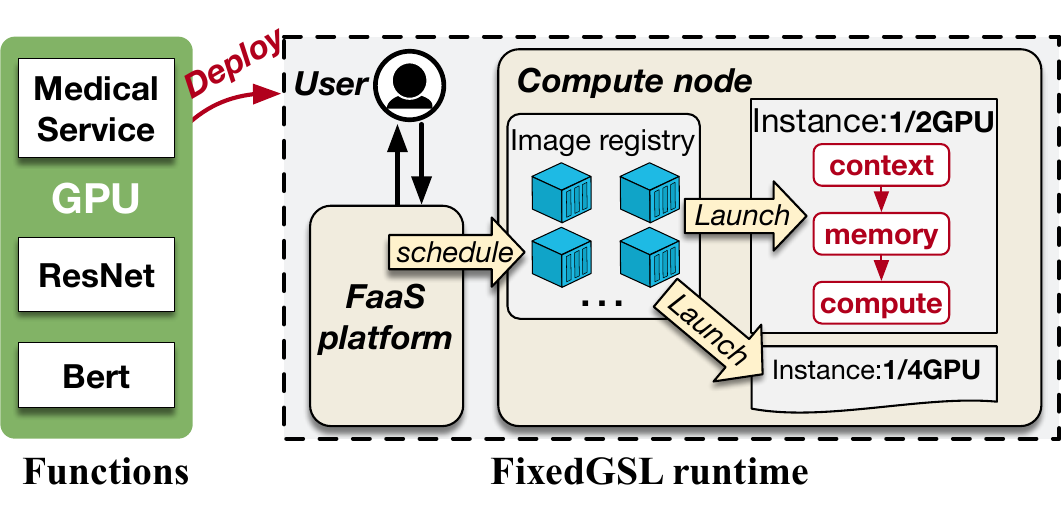}
	\caption{Workflow of \sota{}.}
	\label{fig:instanceGSL}
 \vspace{-2mm}
\end{figure}

Cloud providers have spent some efforts to integrate GPU capabilities into serverless computing platforms for processing compute-intensive applications.
The well-established solutions for GPU-enabled serverless computing are Azure Function~\cite{azurefaasgpu} and Alibaba Function Compute~\cite{aliyun-faas-gpu}. 
Both platforms adopt the way in \autoref{fig:instanceGSL} to schedule GPU functions. 
Specifically, They launch size-fixed container instances with GPU calls wrapped in them to manage GPU functions. 
We refer to this mode as instance-fixed GPU serverless computing (denoted by {\it \sota{}}), as the scheduled instances have predetermined memory and computing specifications. 

Although \sota{} enables GPU in serverless computing, its coarse-grained management brings the {\it long setup time} of GPU functions and {\it low function throughput} on the GPU.


\textbf{Long setup time.} Similar to CPU functions, execution of each GPU function in the \sota{} requires launching a container first. 
However, GPU functions suffer from additional setup time beyond the setup stages experienced by CPU functions. Compared to CPU functions, there are two extra setup stages in GPU functions: {\it GPU context preparation} and {\it GPU data preparation}. 
Our experimental findings, as detailed in \S~\ref{sec:moti_long}, reveal that GPU context preparation requires approximately 280 milliseconds, while GPU data preparation consumes tens to hundreds of milliseconds.

\textbf{Low function throughput.} Our investigation in \S\ref{sec:moti_low} demonstrates that \sota{}'s supported function throughput is only 12.3\% of theoretical peak throughput. There are two main reasons for this gap.
On one hand, while size-fixed memory of \sota{} facilitates memory management for cloud service providers, it also restricts the number of GPU functions residing on the GPU simultaneously.
For instance,  when the memory size of a function instance is set to 2GB, only 20 functions can reside on A100-40GB.
On the other hand, concurrent function invocations all use the same data-loading path (network, disk, and PCIe) for data preparation. The contention of these data-loading paths further degrades the end-to-end latency of each function invocation, thus leading to low throughput.


Similar to the methods for optimizing CPU serverless frameworks~\cite{sand,firecracker,zijunli2022rund}, several intuitive approaches could be used. 
However, these approaches are still far from optimal for GPU serverless.
For instance, preparing the GPU context in advance can alleviate long setup time by eliminating context-creating time. It does not handle the data preparation for launching GPU kernels and introduces additional GPU memory consumption (414MB for each pre-warmed function instance on A100).
Besides, flexible memory allocation could allow more function invocations on a single GPU. Although it
mitigates the insufficient function invocations, the contention of the data-loading path will increase, which further results in poorer throughput.

This paper proposes \textbf{\sysname{}}, the first GPU serverless framework that enables fast setup and high function throughput.
Specifically, each function invocation could be processed with as little setup time as possible.
Meanwhile, \sysname{} reduces the data-loading contention with flexible memory management.
Under these cases, GPU's computing resources are fully utilized, thus resulting in a high system throughput.

To tackle the long setup issue, the key insight of \sysname{} is that \textbf{the massive data loaded through PCIe for a GPU function is knowable in advance of its actual execution.}
The information for the data retrieved from the host side is typically already stored within the metadata of the incoming request.
This implies that the data preparation can be proactively performed for a GPU function without a pre-created GPU context. Based on this insight, \sysname{} proposes \textit{parallelized function setup} mechanism to reduce the setup time by parallelizing the GPU context creation and data preparation.

The parallelized setup mechanism involves an unified memory daemon and a taxon shim.
The memory daemon is designed to proactively load data for GPU functions.
The shim is used for intercepting the GPU calls, and classifying them into memory calls and kernel calls. 
Specifically, the shim communicates with the unified memory daemon to get the actual GPU memory address when intercepting the memory call. Meanwhile, it also ensures that data is indeed ready on GPU when launching GPU kernels.
Therefore, \sysname{} can quickly launch a GPU function without the pre-created GPU context and wasted GPU memory.

In fact, the unified memory daemon already enables flexible memory management for \sysname{}. \sysname{} now only needs to reduce the contention of the data-loading path for high throughput.
We find another key insight that \textbf{appropriate memory sharing can avoid repeated data preparation, thus reducing unnecessary data-loading contention.}
Fortunately, many memory allocations in GPU functions are read-only memory, e.g., the weight of DNN models. Such memory can be safely shared across invocations to the same GPU serverless function.

Based on this insight, we propose sharing-based memory management.
\sysname{} first exposes the interface for users to specify the read-only memory.
Then, it exploits the memory characteristics in two ways to reduce the data preparation of the invocations to the same GPU function.
As for concurrent invocations, read-only memory and GPU context are directly shared.
As for invocations comming adjacently, \sysname{} devises a multi-stage resource exit scheme. When an invocation finishes, its occupied resources are released in multiple stages according to the order of read-only data, GPU context, CPU context, and container. In these stages, as long as a request is coming in, the exit will stop, and the new invocation will reuse the unreleased resources.
Therefore, \sysname{} is able to achieve high throughput with less memory access through the data-loading path. It is worth noting that context sharing here does not reduce data-loading contention but further saves setup time.

The main contributions of this paper are as follows:
\begin{enumerate}[leftmargin=*]
    \item \textbf{Identifying the root causes that lead to long setup and low throughput in existing GPU serverless systems.} Based on the investigations, we propose \sysname{} with appropriate setup boost and fine-grained memory-sharing methods to address the issues.
    \item \textbf{Designing an elegant mechanism that efficiently parallelizes data preparation and context creation.} \sysname{} utilizes it in order to accelerate the setup of GPU functions, when there are no available resources for reuse.
    \item \textbf{Proposing approaches that reduce contention of the data-loading path for maximizing the overall throughput of GPU serverless.} \sysname{} achieves the goal through fine-grained memory sharing, especially the multi-stage resource exit scheme.
\end{enumerate}

We have implemented and deployed \sysname{} on a cluster with A100 GPUs. We also evaluated it using DL tasks~\cite{bert,deepspeech,inception,lstm,nasnet,resnet,seq2seq,vgg} and scientific benchmarks~\cite{parboil}. Our experimental results show that \sysname{} reduces the function duration by $13.3\times$, and improves function density by $1.22\times$ compared with the state-of-art serverless frameworks.


%% file: contents/related.tex
\section{Related Works\label{sec:related}}
There are a few prior works~\cite{orsca, dgsf, molecule, aliyun-faas-gpu, azurefaasgpu} on integrating GPU in serverless computing. 

The usable GPU serverless platforms are available in Alibaba Cloud~\cite{aliyun-faas-gpu} and Microsoft Azure~\cite{azurefaasgpu}. 
For tradeoffs between GPU accessibility and the portability of existing serverless platforms, they schedule the GPU function at the granularity of size-fixed GPU instances (\sota{}). 
ORSCA~\cite{orsca}, DGSF~\cite{dgsf} provided GPU functions on CPU-only nodes with pre-created GPU context on remote GPUs.
It is achieved through API-remoting.
The high memory overhead for pre-warming and high communication overhead make these works not applicable in the production environment.
We compare \sysname{} with \sota{} and DGSF in \S~\ref{sec:eval}.
In Molecule~\cite{molecule},
a shim layer is proposed to unify the usage of various devices such as DPU, GPUs, and other accelerators in serverless computing.
Molecule is orthogonal to \sysname{}'s goal, and they can be combined together for optimizations.

Some prior works support fine-grained memory and computation management through hijacking driver-level API,
such as Alibaba cGPU~\cite{cgpu}, Tencent qGPU~\cite{tencentQgpu}, vCUDA~\cite{shi2011vcuda}, qCUDA~\cite{lin2019qcuda} and GPUshare~\cite{goswami2016gpushare}.
These technologies provide isolated GPU memory and computing capacity.
NVIDIA also has its own GPU fine-grained resource-sharing solutions, such as MPS~\cite{MPS}.
There are also some works that have proposed GPU-sharing solutions in fixed scenarios, such as deep learning applications~\cite{xiao2020antman, bai2020pipeswitch, yu2020fine}.
These techniques are not aware of the characteristics of GPU serverless and do not help accelerate the setup and increase the function density for existing GPU serverless frameworks.
However, they can be integrated into \sysname{} to provide the mechanism of better memory and computation isolation.

There are also many prior works on optimizing the startup time in serverless computing~\cite{li2022help, oakes2018sock, ustiugov2021benchmarking, van2017spec, vrable2005scalability, mohan2019agile}.
Replayable Execution~\cite{replayable}, Firecracker~\cite{firecracker}, and Catalyzer~\cite{catalyzer} are snapshot and fork-based optimizations. SAND~\cite{sand} used a multi-level sandboxing mechanism to improve the performance of the application. These works only accelerate the host-side operations, making GPU functions still suffer from long setup and extra memory consumption.

%% file: contents/motivation.tex
\section{Background and Motivation\label{sec:moti}}

In this section, we first present state-of-the-art GPU serverless frameworks and the used benchmarks. Then, we analyze the inefficiencies of existing serverless frameworks and the reasons behind them. 

\subsection{GPU Serverless and Benchmarks}\label{sec:benchmarks}

\begin{table}
\caption{The benchmarks used for investigation.}
\label{tb:benchmarks}
\centering
\small
\begin{tabular}{c|c}
\hline
\textbf{Task}                                                                                    & \textbf{Task Type}          \\ \hline\hline
\begin{tabular}[c]{@{}c@{}}vgg11~\cite{vgg}, inception3~\cite{inception},\\ resnet50~\cite{resnet}, nasnet~\cite{nasnet}\end{tabular} &
  Computer Vision \\ \hline
deepspeech~\cite{deepspeech}, seq2seq~\cite{seq2seq} & Speech Recognition  \\ \hline
bert~\cite{bert}   & Natural Language Processing \\ \hline
lbm, mrif, tpacf~\cite{parboil}   & Scientific Computing   \\ \hline
\end{tabular}
\end{table}

GPU container toolkit~\cite{gpu-container-toolkit} is often used to integrate GPU into existing serverless frameworks seamlessly. The GPU calls of functions are forwarded to the underlying GPU driver by the GPU container toolkit. The container toolkit also provides the memory and computing resource partition to meet the multi-tenant requirements of GPUs in serverless computing. Both GPU-enabled Azure Functions~\cite{azurefaasgpu} and Alibaba Function Compute~\cite{aliyun-faas-gpu} use the toolkit. With the help of the GPU container toolkit, they are designed with the \sota{} rules for the convenience of management. That is, each GPU function is wrapped into the container equipped with GPU instances. These GPU instances have limited specifications, and the memory allocation granularity is 1 Gigabyte.

Considering that GPU serverless is suitable for a variety of fields, we also use multiple scientific computing tasks as the benchmarks in addition to the DL inference tasks. We use seven DNN models concerned by the DL inference system~\cite{ma2020rammer} and three scientific tasks from Parboil~\cite{parboil} as our GPU function benchmarks. \autoref{tb:benchmarks} presents the benchmark details. The DL models cover computer vision, speech recognition, and natural language processing. The scientific computing tasks include fluid mechanics ($lbm$), medical assistance ($mrif$), and astrophysics ($tpacf$). 

We run the benchmarks on a node with one NVIDIA A100 (40GB memory) GPU and use \sota{} to manage the GPU resources. Upon each new request, \sota{} allocates memory for the function in 1 Gigabyte granularity. Specifically, the function's memory usage is rounded up from the real memory usage. The detailed hardware and software configurations are described in \autoref{tb:setup} of \S~\ref{sec:eval_setup}. 

\subsection{Inefficiency of \sota{}}

In this subsection, we measure the response latencies of the benchmarks, and the system throughput with \sota{}. The experiments show that \sota{} results in long response latency and low system throughput. The long response latency comes from the extra setup stages for preparing the GPU functions. The low function density is attributed to the resource contention in the data preparation stages.

\subsubsection{Long Setup Time}\label{sec:moti_long}

\begin{figure}
\centering
\includegraphics[width=\columnwidth]{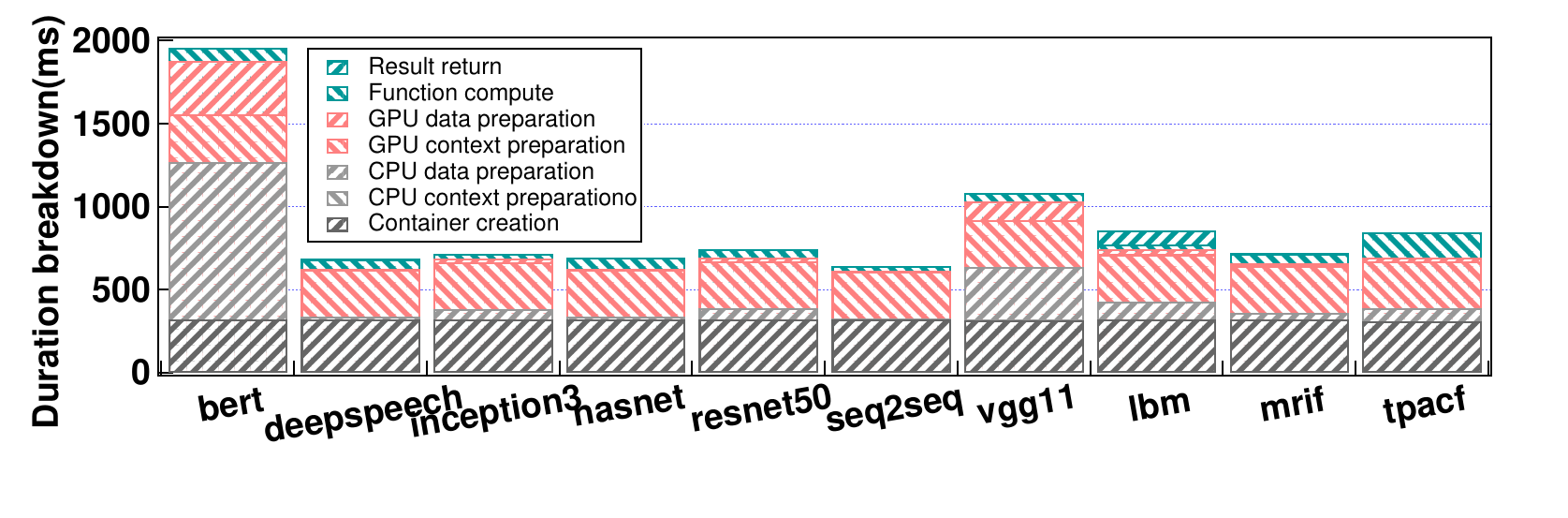}
\caption{The duration breakdown of all GPU functions.}
\label{fig:moti_duration}
\end{figure}

We first measure the end-to-end duration of a GPU function using \sota{}, and breakdown the duration. The function invocations are generated in a close-loop manner, eliminating the impact of queuing and resource contention.

\autoref{fig:moti_duration} shows the duration breakdown of all the benchmarks with \sota{}. The end-to-end duration for GPU functions can be segmented into eight stages, including container creation, CPU context preparation, CPU data preparation, GPU context preparation, GPU data preparation, function computation, and result return. Both data preparation stages incorporate memory allocation and corresponding data transfer. For CPU data preparation, GPU functions load data from the database through Disk or Network. For GPU data preparation, GPU functions transfer data from CPU memory to GPU memory through PCIe. We refer to the five stages before computation as setup stages.

The experimental results reveal that the computation time only accounts for $7.1\%$ of the end-to-end duration on average, with a maximum of $17.8\%$. GPU functions are suffering from long setup time. Among these setups, container creation is the most time-consuming ($39.6\%$ of end-to-end latency on average). As for the container creation, we have two observations here. First, there is no difference in startup time between GPU container and CPU container. Secondly, the calling commands of the two containers are very similar. GPU container creation time can be optimized using the pre-warm techniques proposed for CPU container~\cite{aliyun-faas-gpu, li2022help, oakes2018sock}.

When the container creation time is optimized away, the overall time of all functions is reduced. In this case, the computation time takes up 12.1\% of the overall time, and the preparation time takes up 86.3\%. Among these setups, GPU context preparation and data preparation (CPU/GPU data loading) consume the most time (86.1\% of the overall time). Because GPU context is created implicitly and GPU data loading relies on created GPU context, GPU functions are executing these GPU-related setups in serial. Therefore, GPU functions are suffering long setup time.

\subsubsection{Low System Throughput}\label{sec:moti_low}

This experiment reports the achievable system throughput of all benchmarks with \sota{}. In this experiment, requests are generated with a Poisson distribution and are scheduled in an open-loop manner. Such experiment configuration can maximize the actual loads to the GPU. In order to obtain the peak system throughput, we gradually increase the load until \sota{} fails to process the received requests stably.

The throughput of each application over a period $T_{period}$ using FixedGSL is also collected and denoted as $Throughput_{app}$. We also calculate the theoretical throughput $Throughput_{theo}$ in \autoref{eq:throughput-theo}, which simply considers the application's computation time $T_{comp}$. Based on the $Throughput_{theo}$, the normalized throughput performance $Throughput_{perf}$ of each application is further calculated.
\begin{equation}
\label{eq:throughput-theo}
\begin{aligned}
Throughput_{theo} &= T_{period} / T_{comp} \\
Throughput_{perf} &= Throughput_{app} / Throughput_{theo} 
\end{aligned}
\end{equation}

\begin{figure}
\centering
\includegraphics[width=\columnwidth]{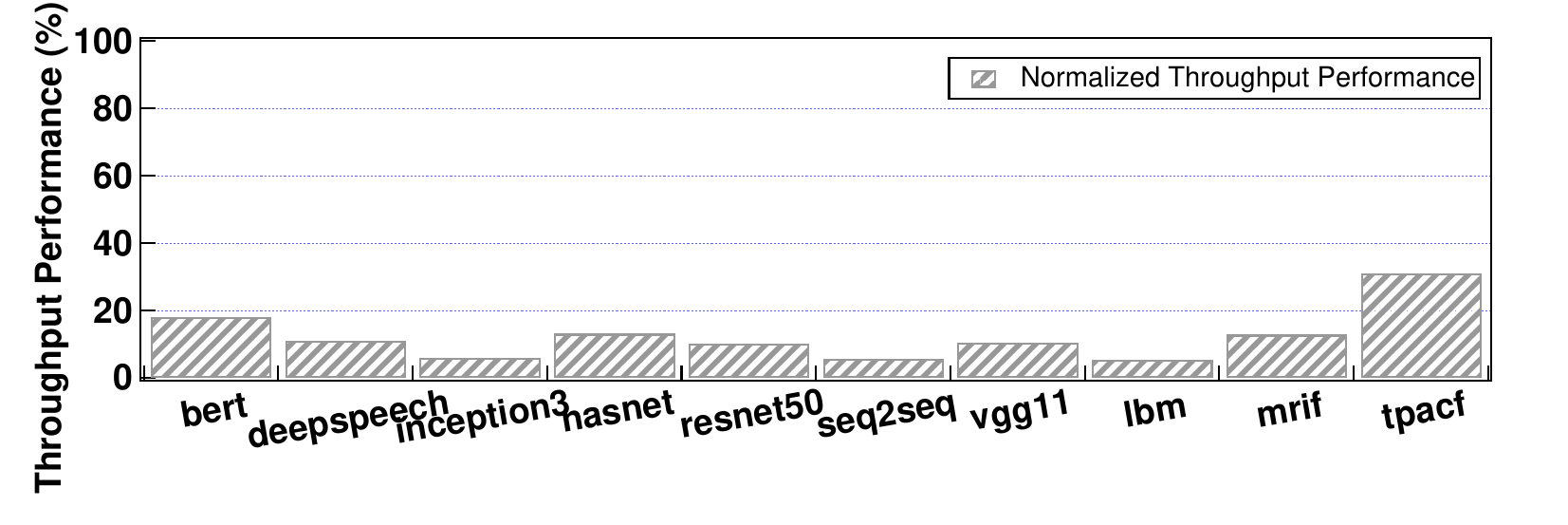}
\caption{The system throughput of different functions under \sota{}.}
\label{fig:moti_fixedgsl}
\end{figure}

\autoref{fig:moti_fixedgsl} shows the normalized throughput performance of each application. On average, the actual throughput is only 12.3\% of its theoretical value. One straightforward reason for the poor performance is the fixed memory allocation, which restricts GPU from serving more application queries. Furthermore, we find that another reason for the low throughput is the resource contention in the data loading paths.


\begin{figure}
\centering
\includegraphics[width=\columnwidth]{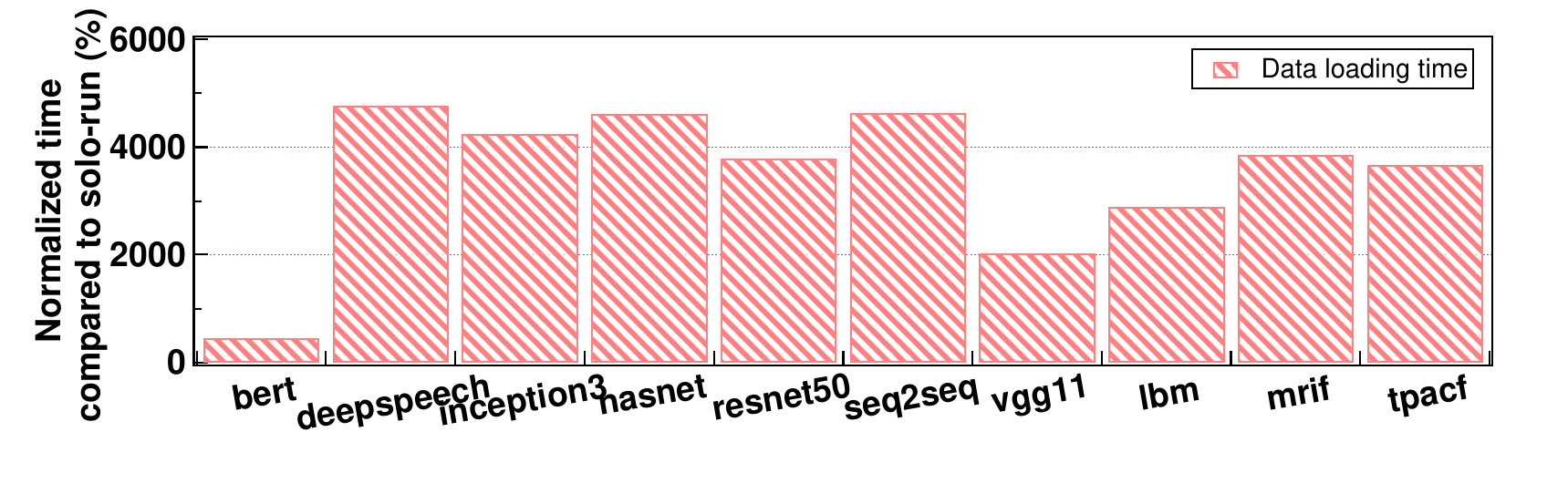}
\caption{The average setup time of all applications normalized to the solo-run case under \sota{} .}
\label{fig:moti_setup}
\end{figure}

\autoref{fig:moti_setup} shows the average data loading time of all applications under \sota{}. This time is normalized to the data loading time without any resource contention in \autoref{fig:moti_duration}. As shown in the figure, each application suffers 34.9$\times$ data loading time under \sota{}. Specifically, loading data to CPU memory from database and copying data to GPU memory through PCIe encounter severe resource contention. As a consequence, \sota{} has the low throughput problem. 

In summary, \sota{} encounters long setup time and low system throughput. \textbf{The long setup time comes from the extra setup stages of the GPU function compared to the CPU function. The low system throughput comes with resource contention in the data loading paths.}

%% file: contents/design.tex
\section{Design Overview}
\label{sec:overview}

Faced with the above problems, we propose \sysname{}, a GPU serverless framework that enables fast function setup and high function throughput. In this section, we illustrate the architecture of \sysname{} and corresponding programming model. 




\subsection{SAGE Architecture}

\begin{figure}
\centering
\includegraphics[width=\columnwidth]{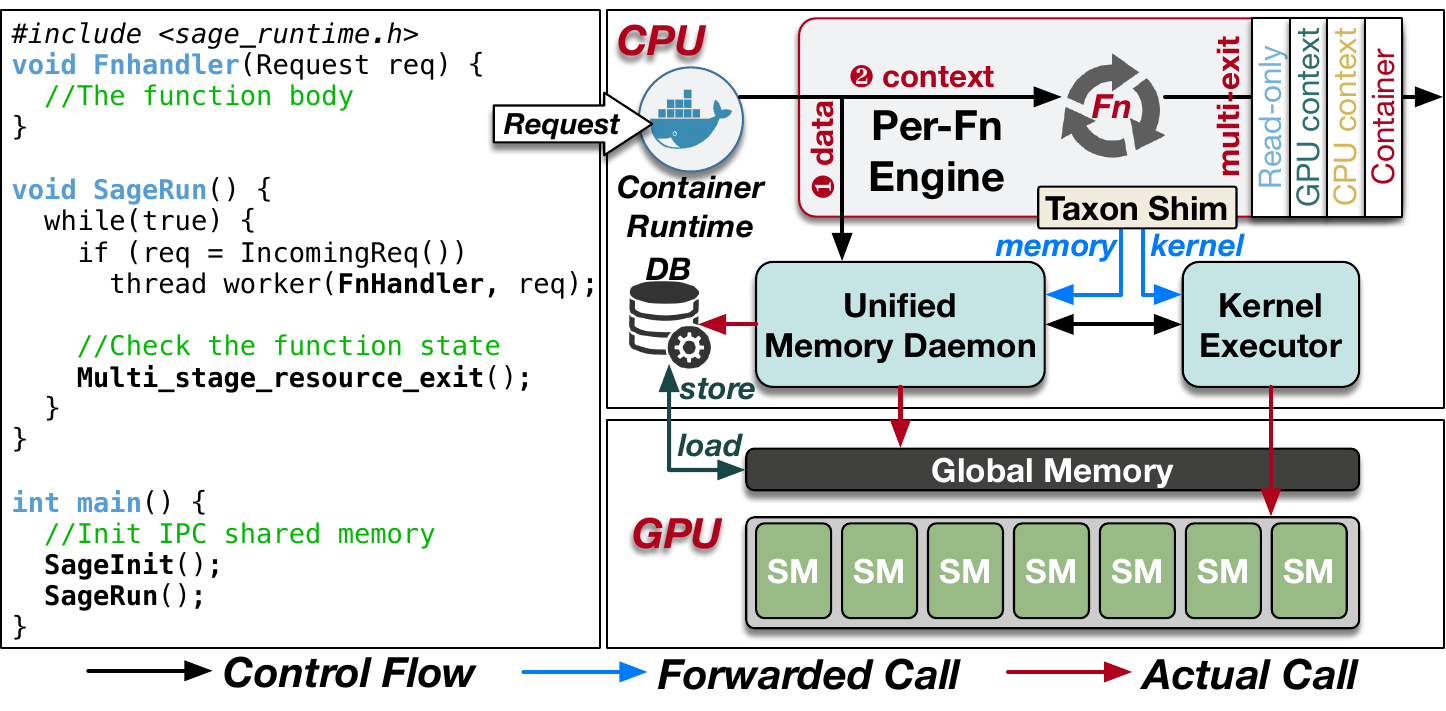}
\caption{The overview of SAGE.}
\label{fig:overview}
\vspace{-2mm}
\end{figure}

\autoref{fig:overview} shows the overview of \sysname{} and its corresponding programming interface. As shown in the right part of \autoref{fig:overview}, \sysname{} comprises four modules: the \textit{per-function engine}, the \textit{taxon shim}, the \textit{unified memory daemon}, and the \textit{kernel executor}. \sysname{} proposes two innovative mechanisms to enhance the function latency and throughput performance for the GPU serverless system, which are parallelized function setup and sharing-based memory management.

When the upper cluster scheduler assigns a GPU function request to \sysname{} runtime, \sysname{} first starts a container with GPU capability. Inside the container, a per-function engine is then initialized. The engine first extracts the data that can be loaded in advance and notifies the memory daemon to perform data preparation. After that, the engine continues to trigger the execution of the function. Because the memory daemon works independently, the data preparation and the context preparation inside the engine are parallelized.

When the function handler is being executed, taxon shim intercepts all GPU calls and dispatches them respectively. Specifically, if it is a memory-related call, the shim forwards it to the memory daemon. The memory daemon returns the actual device memory address of preloaded data to the function handler. GPU Kernel-emitting calls are forwarded by the shim to the kernel executor. The kernel executor then communicates with the memory daemon to ensure that related GPU memory is correctly prepared.

\sysname{} deploys only one memory daemon on a single GPU. The memory daemon and the function engine collaboratively support the sharing-based memory management mechanism. When multiple invocations of the same function are scheduled to the same GPU, the read-only data will be only loaded upon the first invocation. The other invocations could share the same copy, which reduces the data loading requirement and the resource contention in the data loading paths. As a consequence, the function throughput is improved. 

In addition, the per-function engine is also shared with the latter invocations. \sysname{} keeps the engine alive within a period. To enhance the sharing efficiency, the lifecycle of the launched per-function engine ends with a multi-stage exit method, rather than terminating directly upon the completion of the invocation's execution. Specifically, the occupied resources of a function engine are released in multiple stages according to the order of read-only data, GPU context, CPU context, and container.


\subsection{Programming Model}
The left part of \autoref{fig:overview} presents an example of \sysname{}'s programming interface. Our current implementation of \sysname{} only supports \texttt{C++} language runtime. It should be trivial to provide other language bindings (e.g., \texttt{Python}), as GPU relies on \texttt{C++} to provide its functionality. The provided interfaces follow the design pattern of the existing serverless framework AWS Lambda~\cite{aws-cplus} that supports \texttt{C++} runtime. 

The developer wraps the GPU function code within the function handler with a pre-defined signature. \sysname{} also requires developers to store metadata of the data loaded from external sources when defining \texttt{Request} struct, which is detailed in \S~\ref{sec:parallel}. The read-write attribute of data also needs to be designated to ascertain whether it can be shared as read-only data. \sysname{}'s runtime can then identify them for setup parallelization and memory sharing.

\texttt{SageInit()} and \texttt{SageRun()} are two APIs provided to initialize the \sysname{}'s runtime and invoke the function execution upon an incoming request. These two APIs do not need user definition. 
Later, we will present the interface details about how \sysname{} parallelizes the context creation and data preparation. Besides, we will also give an example of how to utilize our interface to transform the original GPU application into a GPU function for \sysname{}. 


%% file: contents/manager.tex
\section{Parallelized Function Setup}\label{sec:manager}
In this section, we present our methods for optimizing function setup, while the function is first invocated on the GPU. These methods are crucial in accelerating the function setup process and the end-to-end latency.

\subsection{Parallelizing data and context} \label{sec:parallel}

\begin{figure}
\centering
\includegraphics[width=.95\columnwidth]{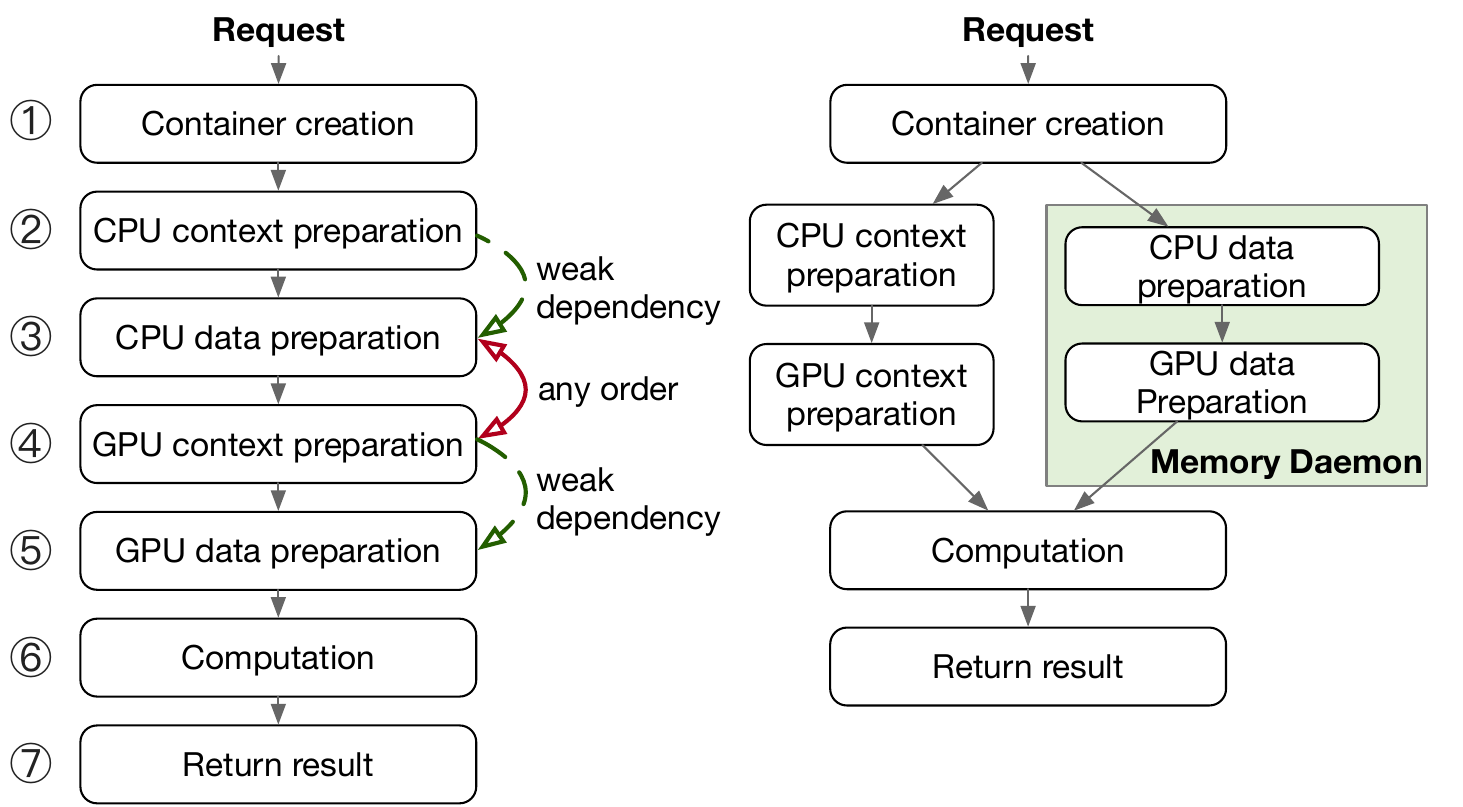}
\caption{The GPU function's simplified workflow.}
\label{fig:workflow}
\end{figure}

We outline the simplified execution flow of the GPU functions in \autoref{fig:workflow}, which consists of seven stages.
(1) The GPU scheduler creates a container for the GPU function.
(2) The GPU function establishes its own CPU context and corresponding function engine.
(3) The GPU function requests CPU-side memory and loads data from a designated database based on the invocation request.
(4) The GPU function generates its own GPU context and initializes the corresponding library.
(5) The GPU function requests GPU memory and transfers the data from CPU to GPU.
(6) The GPU function starts computing.
(7) Upon completion of computing, the GPU function returns the result.

Among these stages, stage-3 and stage-4 could be executed in any order for no dependency. Meanwhile, stage-5 and stage-6 can also occur in any order because the GPU memory allocation and the GPU kernel launch could happen at any time. However, developers usually prioritize preparing the data first before launching a series of GPU kernels. In this way, kernel launching overhead is hidden, and GPU utilization is improved. This means that the first five stages are executed sequentially prior to the computation stage. As a result, the GPU function experiences a long setup.

\begin{figure*}
\centering
\includegraphics[width=2\columnwidth]{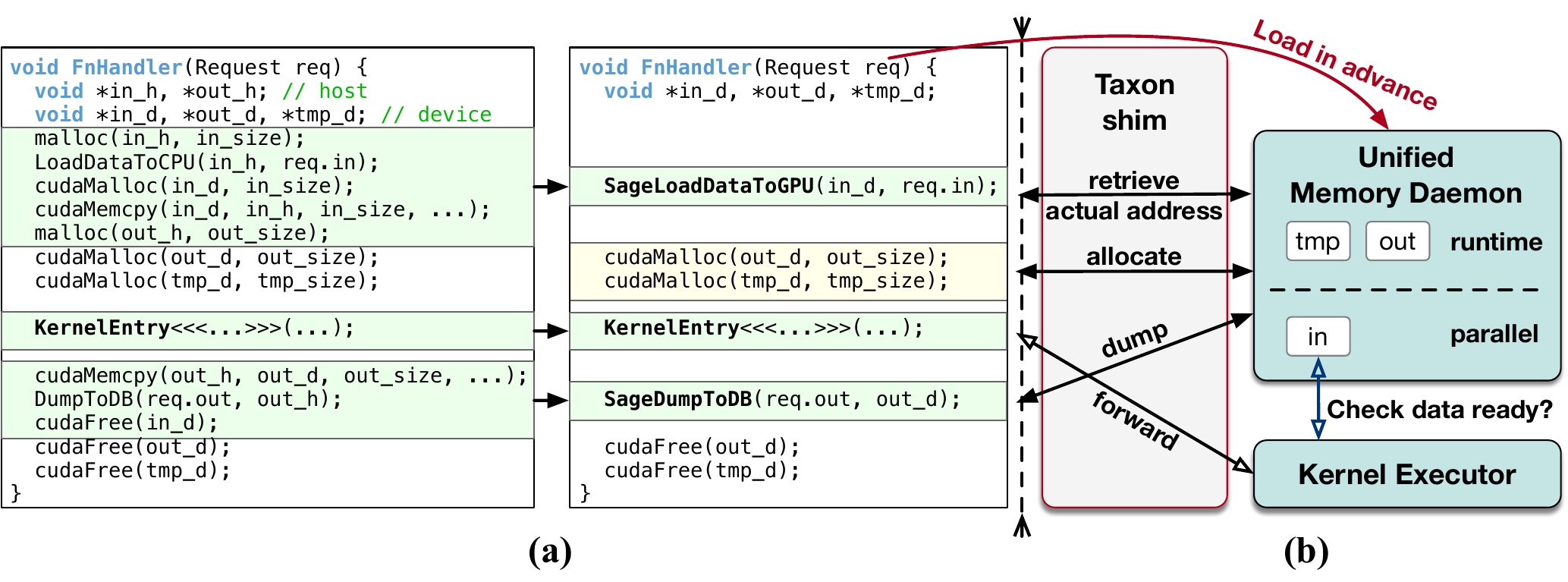}

\caption{The figure-(a) presents an example of converting existing GPU application to a \sysname{} supported GPU function with provided programming interface; The figure-(b) demonstrates the working mechanism of taxon shim, including intercepting GPU calls and re-dispatching them.}
\label{fig:shim}
\end{figure*}

\begin{figure}
\centering
\includegraphics[width=.95\columnwidth]{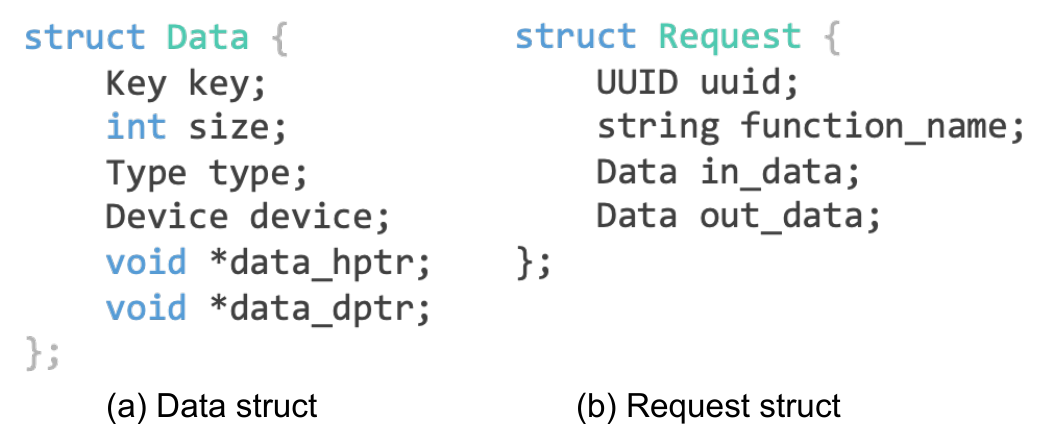}
\caption{Programming interface of \sysname{}.}
\label{fig:programming}
\vspace{-8mm}
\end{figure}


After a deeper analysis of \autoref{fig:workflow}, we have two observations.
First, GPU functions always load data from external sources. For instance, DL inference services commonly read model inputs and weights from databases, while scientific computing services require input data or updated data to be sourced from databases.

Secondly, both CPU and GPU support inter-process communication. CPU and GPU could use data from other processes for computation. Corresponding experimental results show that a GPU function can compute correct results through the exclusive use of data from other processes. This means that context and data are not firmly intertwined. 

Based on these two observations, we have a key insight: \textbf{GPU data preparation can be delegated to another process, enabling parallel preparation of both the GPU data and GPU context.} Further, we reconsider the execution flow of a GPU function and have three findings. (1) Container creation, CPU context creation, and GPU context creation must be performed in a strict sequence. (2) There is no strong bond between the GPU data and GPU context, and GPU data loading depends on CPU data loading. (3) GPU computation depends on both GPU context and GPU data.

We reconstruct the dependency graph for the GPU function setup stages. As shown in \autoref{fig:workflow}, we could divide the function setup stages into two distinct parts, namely GPU context preparation and GPU data preparation. The function engine is designated with GPU context preparation, whereas the memory daemon handles GPU data preparation from external sources. They are executed in parallel.

To implement the aforementioned process, the function engine must communicate the data-loading information with the memory daemon ahead of actual execution.
To this end, we devise a data structure to store the positional details of the external data required by a request, as illustrated in \autoref{fig:programming}. 

Specifically, \texttt{Request} contains all the data that require loading from the databases. The database information is stored in the \texttt{key} of the \texttt{Data} structure. When the memory daemon receives a request from the function engine, it immediately initiates the data-loading process. After the data loading is finished, the  \texttt{data\_dptr} could used to perform the computation. Leveraging this approach, GPU context preparation and GPU data preparation can occur concurrently, thereby enabling the fast function setup.

\subsection{Supporting Prallelization with Taxon Shim}\label{sec:shim}

Loading data proactively through the unified memory daemon can invalidate the execution of the original GPU program.
Previously the program interacts with the GPU directly, the GPU's built-in scheduler ensures the correctness of memory operation, kernel launching, etc.
With the unified memory daemon, there has to be a consolidated design to achieve exactly the same functionality.
Then, the taxon shim is designed in response to the particular need.

\subsubsection{Programming of Function Handle}
The taxon shim first provides APIs to access the address of the previously prepared memory by the memory daemon.
For loading data from external resources, \sysname{} designs an interface termed \texttt{SageLoadToGPU()}, which returns the actual memory address.
For saving data to external destinations, \sysname{} also designs a similar interface called \texttt{SageDumpToDB()}.
Developers use these two APIs for loading and returning data inside the function handler.
When it comes to common GPU function calls like \texttt{cudaMalloc()} and kernel launching, the taxon shim intercepts both of them.
The difference is that a memory call will be forwarded to the daemon, while a kernel call will be forwarded to the kernel executor.

\autoref{fig:shim}-(a) depicts an illustrative example of using the programming interface.
As shown in the figure, a function handler may utilize either \texttt{SageLoadToGPU()} for data loading or \texttt{cudaMalloc()} for straightforward allocation.
When the GPU function calls \texttt{SageLoadToGPU()}, the memory daemon returns a pointer to the function.
Notably, \texttt{SageLoadToGPU()} is an asynchronous operation, and the data loading may not finish after invoking it.
Kernel entry is then also intercepted.
At the end of the handler, \texttt{SageDumpToDB()} returns the result to the memory daemon, which is responsible for dumping the data into the database.

\subsubsection{Ensuring Function Correctness}
\autoref{fig:shim}-(b) demonstrates how taxon shim interacts with memory daemon and kernel executor. 
It is easy to see that the taxon shim forwards all GPU calls based on their categories.
All memory-related calls are forwarded to the memory daemon, and all kernel calls are forwarded to the kernel executor.
This is because we need a central memory coordinator to ensure that the pre-loaded data is correctly prepared.

As the taxon shim forwards the GPU calls one by one, the memory daemon also processes the memory calls one by one.
When the kernel executor receives the kernel, it only needs to communicate with the memory daemon to verify whether all the data required by the kernel has been prepared. If all data is ready, the kernel executor can directly launch the kernel to the GPU.
Based on these processes, the taxon shim could support the parallelized cold setup without any faults.

%% file: contents/allocator.tex
\section{Sharing-based Memory Management}\label{sec:allocator}

Following optimizing the function setup time, it is imperative that we further reduce the resource contention in the data loading paths and address the low function throughput attributed to existing GPU serverless solutions. In this section, we present our sharing-based memory management method, which solves the above issue.

\subsection{Memory Usage Analysis} \label{sec:analysis}

To reduce resource contention in the data loading paths, we initially conduct experiments to investigate the memory usage of mainstream GPU functions. Our investigation reveals that GPU function memory usage can be divided into three categories: context memory, read-only memory, and writeable memory. Context memory is implicitly allocated during the initial execution of the GPU application and is utilized to store the runtime data. Read-only memory represents the read-only data of GPU function execution. Writeable memory, on the other hand, denotes the writable data of GPU function execution.

\begin{table}
\caption{The memory usage characteristics of GPU functions.}
\label{tb:memory}
\centering
\scriptsize
\begin{tabular}{c|c|c|c|c|c}
\hline
\textbf{benchmark} & \textbf{context} & \textbf{read-only} & \textbf{writable} & \textbf{explicit} & \textbf{\begin{tabular}[c]{@{}c@{}}read-only \\ ratio(\%)\end{tabular}} \\ \hline
\hline
bert       & 414   & 1282.5 & 60.1  & 1342,6 & 95.5    \\ \hline
deepspeech & 414   & 24.8   & 6.9   & 31.7   & 78.8    \\ \hline
inception3 & 414   & 91.1   & 11.7  & 102.8  & 88.6    \\ \hline
nasnet     & 414   & 20.3   & 11.8  & 32.1   & 63.5    \\ \hline
resnet50   & 414   & 97.7   & 11.9  & 109.6  & 89.2    \\ \hline
seq2seq    & 414   & 6.1    & 0.1   & 6.2    & 98.9    \\ \hline
vgg11      & 414   & 506.8  & 38.0  & 544.8  & 93.0    \\ \hline
lbm        & 414   & 0      & 330   & 330    & 0       \\ \hline
mrif       & 414   & 0      & 22    & 22     & 0       \\ \hline
tpacf      & 414   & 0.1    & 28.3  & 28.3   & 0.4     \\ \hline
\end{tabular}
\end{table}

\autoref{tb:memory} demonstrates the memory usage allocation of all GPU functions. These functions are implemented by Rammer~\cite{ma2020rammer} and Parboil~\cite{parboil}. As observed from the table, context memory occupies the maximum memory usage of all functions except Bert, accounting for an average of 79.0\% of the overall memory usage. Although context memory occupies a large amount of memory, corresponding experimental results show that the context creation time does not change when multiple function invocations create their contexts. This implies that context creation for function invocations does not interfere with each other.

In addition to context memory, explicit memory accounts for 21.0\% of the overall memory usage. As demonstrated in \S~\ref{sec:moti_low}, concurrent function invocations encounter severe resource contention during data loading. The benchmark functions suffer 34.9$\times$ data loading time compared to the solo-run case. Meanwhile, we observe from this table that read-only memory may occupy more than 90\% of explicit memory usage. Optimizing the data loading process for read-only memory can substantially mitigate memory contention issues in data loading paths.

\subsection{Read-only Memory Sharing}

Existing GPUs only offer basic data transfer APIs through PCIe, which could only handle data transfer tasks sequentially. Meanwhile, the disk and network bandwidth used by CPU loading are also limited. There lack sophisticated interfaces to manage the data transmission of multiple invocations to maximize the efficiency of data loading. Under these circumstances, we find that multiple invocations of the same function can safely share the read-only memory.

While read-only memory remains unmodified during the function execution, the read-only memory sharing between multiple invocations does not cause correctness issues or performance problems. It should be noted that multiple function invocations perform the same computation, which comes from the same service provider. There will also be no malicious memory attack on the GPU. In this case, read-only memory sharing can effectively reduce data loading, which further reduces the resource contention in the data loading paths. Therefore, the overall throughput can be improved.

To support the read-only memory sharing between multiple invocations, the memory daemon needs to identify read-only data in the function. Since function developers could easily distinguish read-only data from writable data, we require the function developers to indicate the data attributes in the data structure shown in \autoref{fig:request}. If a data is read-only, the \texttt{Type} in the \texttt{Data} structure should be set to \texttt{ReadOnly}. Otherwise, it should be set to \texttt{Writable}.

It is worth noting that read-only memory is not limited to the weight of ML inference tasks. The $tpacf$ benchmark in astrophysics also has read-only memory. Any read-only data in the GPU function could be shared by multiple invocations.

\subsection{Multi-stage Resource Exit} \label{sec:stage}

Another commonly used sharing method involves reusing launched containers and keeping them warm for periods of time. Although the initial function call must complete all necessary prerequisite setups (referred to as a cold start), subsequent calls may be processed by the warm container, thereby mitigating undesired startup latency. Leading serverless platforms, including OpenWhisk~\cite{openwhisk} and AWS Lambda~\cite{lambda}, mostly offer support for container reuse.

The warmed container exempts the invocation from some setup stages, which include container creation, context creation, and data loading. However, simply applying the warm strategy to GPU functions may not be a good choice. Specifically, maintaining a warm container for extended periods requires extra memory usage, such as GPU context memory and read-only memory. While GPU memory is scarce, maintaining a warm container for GPU function may cause a decrease in function density.

\begin{figure}
\centering
\includegraphics[width=\columnwidth]{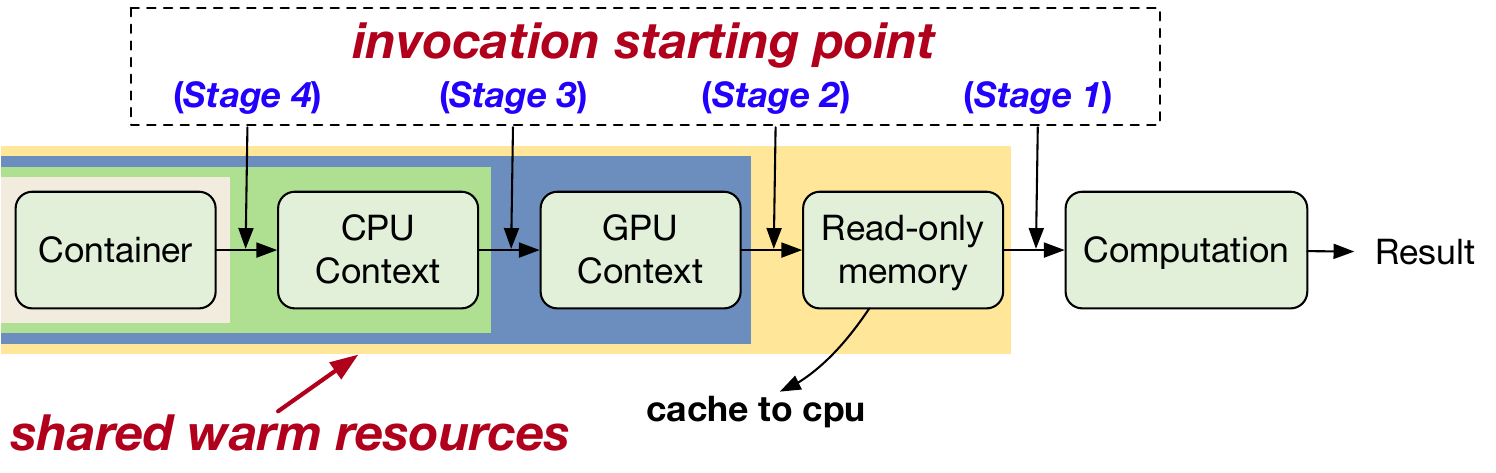}
\caption{The multi-stage resource exit mechanism. The shared resources are marked with different background colors for different stages.}
\label{fig:multi-stage}
\end{figure}

To solve this problem, we leverage the CPU-side context as a suitable intermediary layer for GPU functions. Given that multiple prerequisite stages are required for the computation of the GPU function, we design a multi-stage resource exit mechanism. \autoref{fig:multi-stage} shows our multi-stage exit mechanism.

In the first stage, the GPU function retains all GPU context and read-only data. In the second stage, the GPU function preserves the GPU context, yet caches the read-only data to CPU memory.
This is because CPU memory typically boasts larger storage capacity, lower cost, and faster PCIe transfer times. In the third stage, the GPU function discards the GPU context but still retains the CPU memory and CPU context.
In the fourth stage, the GPU function discards the CPU memory and CPU context but retains the container to minimize the start time of subsequent requests.
Finally, after the container has served its purpose, it is destroyed. 

Leveraging the multi-stage exit mechanism outlined above, we are able to attain a more flexible management of the tradeoff between GPU resource cost and warm containers.
Moreover, it is noteworthy that the time interval of each stage is adjustable.
For instance, configuring each stage's interval to the previous time interval would enable the GPU function to maintain a longer warm state at a lower cost.
Alternatively, configuring the overall duration of the three stages to the previous time interval would allow the serverless platform to support a similar warm effect at a lesser cost. In this paper, \sysname{} configures each stage's interval to the previous time interval, which is set as 30 seconds\cite{li2022help,zijunli2022rund}. 

\subsection{Context Memory Sharing} \label{sec:context}

Note that, while context memory sharing does not mitigate the resource contention in the data loading paths, it still brings considerable context creation time and memory usage. Therefore, we also support the context memory sharing between multiple invocations of the same function. Using the context memory sharing, the function invocations could benefit from the shorter function setup and less memory usage. The shorter function setup further improves the system throughput, and the less memory usage enables the GPU to keep more functions warm on the GPU. Similarly, the invocations of the same function perform the same computation, there will be no malicious attack about the context.

%% file: contents/evaluation.tex
\begin{figure*}
	\centering
	\includegraphics[width=2\columnwidth]{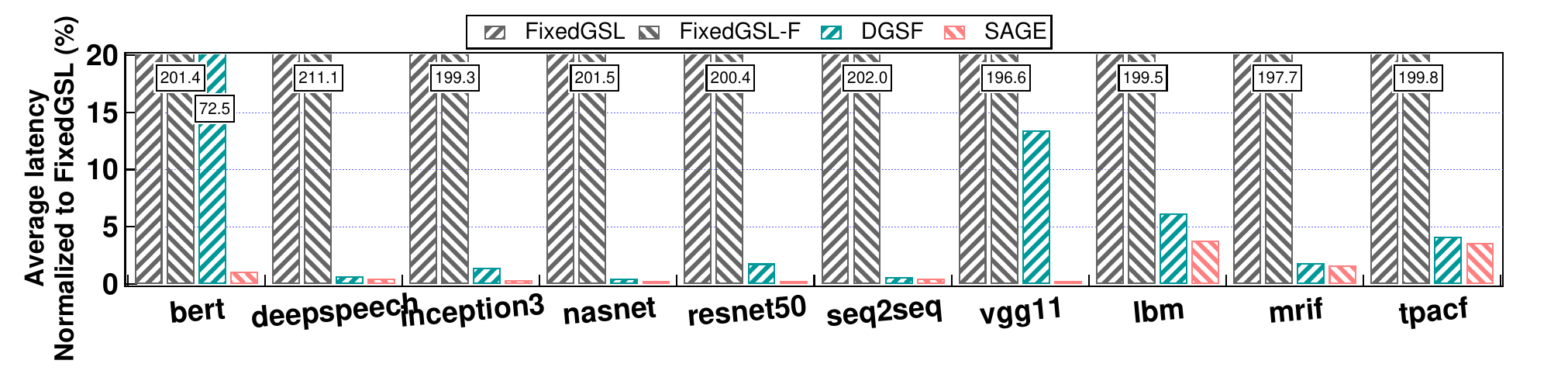}
    \vspace{-6mm}
	\caption{The normalized GPU function's average latency with \sota{}, \sota{}-F, DGSF, and \sysname{} compared with \sota{}. While the \sota{} is the baseline, all the latency results (the gray bar on the far left) with \sota{} is 100. }
	\label{fig:exp-1}
\end{figure*}

\section{Evaluation\label{sec:eval}}


\subsection{Implementaion and Experimental Setup}\label{sec:eval_setup}

\begin{table}
\renewcommand{\arraystretch}{1.2}
\caption{Hardware and software setups in the experiment.}
\label{tb:setup}
\centering
\scriptsize
\begin{tabular}{cl|cl}
\hline
\multicolumn{2}{c|}{}      & \multicolumn{2}{c}{Configuration}            \\ \hline
\multicolumn{2}{c|}{CPU}             & \begin{tabular}[c]{@{}c@{}} Intel(R) Xeon(R) Silver 4216 CPU @ 2.10GHz,\\ Disk: HDD 2T, Cores: 64, DRAM: 125GB  \end{tabular} \\ \hline
\multicolumn{2}{c|}{GPU}             & \multicolumn{2}{c}{\begin{tabular}[c]{@{}c@{}}NVIDIA A100, 40GB HBM\end{tabular}} \\ \hline
\multicolumn{2}{c|}{Software}             & \multicolumn{2}{c}{\begin{tabular}[c]{@{}c@{}}Operating system: Ubuntu 20.04 with kernel 5.4.0\\ Docker version: 20.10.21, NVIDIA driver: 470.161.03\end{tabular}} \\ \hline
\multicolumn{2}{c|}{Container}            & \multicolumn{2}{c}{\begin{tabular}[c]{@{}c@{}}Image: nvcr.io/nvidia/pytorch:22.09-py3\\ CUDA: 11.8, cuDNN: 8.6.0, cuBLAS: 11.11.3\end{tabular}}                    \\ \hline
\end{tabular}
\end{table}

We implement \sysname{} with all components described in \S~\ref{sec:overview}. Our system uses Docker~\cite{nvidia-docker} for the application sandbox, GRPC~\cite{grpc} for the message passing, and MongoDB~\cite{mongodb} for the database. All components of \sysname{} are implemented as independent processes based on C++. 

We use \sota{}, \sota{}-F, and DGSF as the baselines. \sota{} is only capable of allocating memory for functions at a granularity of 1 Gigabyte. We extend \sota{} to \sota{}-F, which supports flexible memory allocation for functions. Additionally, we also choose DGSF, a state-of-the-art serverless system, as another baseline. Each function in DGSF has four pre-created GPU contexts, eliminating GPU context preparation time. Invocations to the same function share these four contexts using a First-Come-First-Serve queue~\cite{dgsf}. Note that, to better distinguish our effect from previous works on the CPU side~\cite{sand,firecracker,zijunli2022rund}, we enhance all the systems with the pre-warmed container. Therefore, there is no container creation overhead for GPU functions.

To evaluate the effectiveness of \sysname{}, we use seven DNN applications from Rammer~\cite{ma2020rammer} and three scientific computing applications from Parboil~\cite{parboil} as our benchmark suite. These applications cover diverse domains such as object recognition, speech generation, natural language processing, fluid mechanics, and real-time medical imaging. Detailed information about the benchmarks can be found in \autoref{tb:benchmarks} of \S~\ref{sec:moti}. Besides, we replay the workload trace from Microsoft Azure Functions (MAF) ~\cite{shahrad2020serverless} for 2 hours in all experiments. After we adapt the benchmarks to GPU functions using our interfaces, we could conduct the experiments. 

The experiments described in \S~\ref{sec:eval-density} to \S~\ref{sec:abalation2} are executed on the Nvidia A100 GPU. All the experiment setups are shown in \autoref{tb:setup}. First, the experiments in \S~\ref{sec:eval-density} demonstrate the function latency performance and the system throughput performance of \sysname{}. Second, the experiments in \S~\ref{sec:exp-moti} are performed to compare with the theoretical performance. Third, the experiments in \S~\ref{sec:more-func} prove the performance of \sysname{} with more functions. Then, we conduct the ablation study for the multi-stage resource exit mechanism and read-only memory sharing scheme in \S~\ref{sec:abalation1} and \S~\ref{sec:abalation2}. Finally, \sysname{} is deployed on a small cluster containing 4 Nvidia A100 GPUs to prove the scalability in \S~\ref{sec:scale_out}.

\subsection{Latency and Throughput}
\label{sec:eval-density}

\begin{figure*}
	\centering
	\includegraphics[width=2\columnwidth]{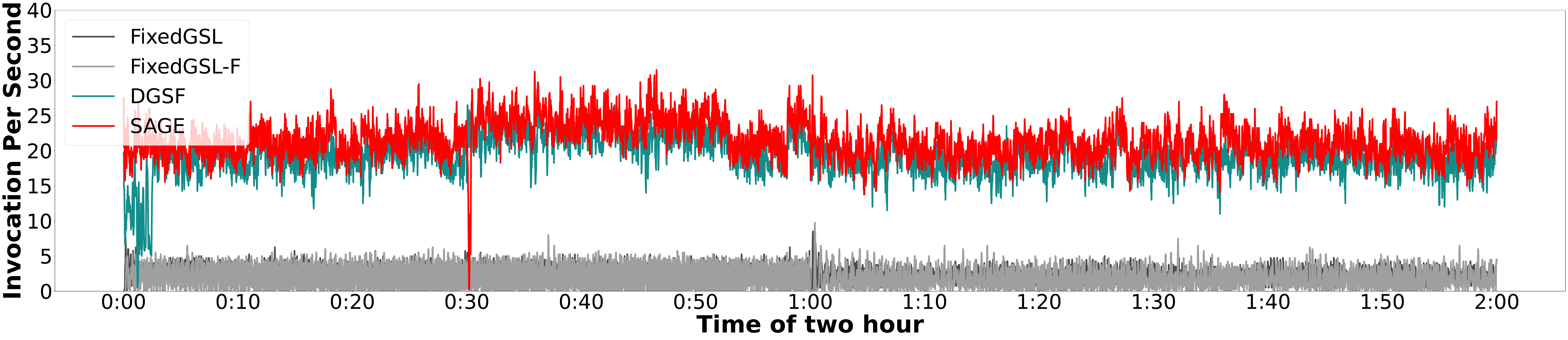}
    \vspace{-2mm}
	\caption{The system throughput (invocations per second) with \sota{}, \sota{}-F, DGSF, and \sysname{}.}
	\label{fig:qps}
\end{figure*}

\begin{figure*}
	\centering
	\includegraphics[width=2\columnwidth]{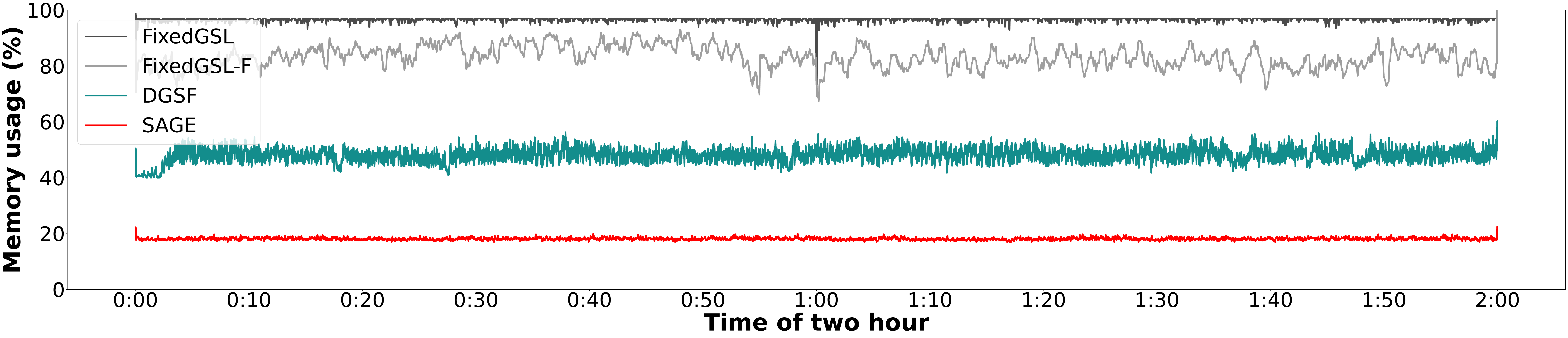}
    \vspace{-2mm}
	\caption{The memory usage ratio with \sota{}, \sota{}-F, DGSF, and \sysname{}.}
	\label{fig:ratio}
\end{figure*}
\autoref{fig:exp-1} illustrates the average latency of ten applications for \sota{}, \sota{}-F, DGSF, and \sysname{}. On average, \sysname{} outperforms \sota{}, \sota{}-F, and DGSF by 193.4$\times$, 391.5$\times$, and 13.3$\times$, respectively. \sysname{} outperforms \sota{}, \sota{}-F, and DGSF on the minimum by 26.3$\times$, 52.4$\times$, and 1.21$\times$, respectively. Besides, we also collect the 99\%-ile latency of ten applications, which are not shown due to page limitation. For the 99\%-ile latency, \sysname{} outperforms \sota{}, \sota{}-F, and DGSF by 54.1$\times$, 109.2$\times$, and 25.4$\times$, respectively.


The improved latency performance of \sysname{} comes from two aspects. Firstly, \sysname{} accelerates the processing of cold invocations by paralleling its context preparation and data preparation.
Secondly, sharing read-only memory and existing context also shortens the end-to-end latency of warm  (invocations to the same GPU function).

As a comparison, \sota{} and \sota{}-F do not prioritize the latency performance of functions. They only schedule the function invocations with the memory usage in a fixed granularity.
Although multiple invocations from the same function in DGSF could enjoy the pre-created GPU contexts, DGSF ignores the resource contention in the data loading path, thus still suffering long data preparation. Our proposed method, \sysname{}, takes both of these factors into account and has been shown to be highly effective in reducing latency.


 

\autoref{fig:qps} illustrates the system throughput of \sota{}, \sota{}-F, DGSF, and \sysname{} during this period. \sysname{}'s system throughput outperforms \sota{}, \sota{}-F, and DGSF by 8.9$\times$, 10.3$\times$, and 1.22$\times$, respectively. The throughput improvement comes from the memory sharing of \sysname{}. The read-only memory sharing between multiple invocations could mitigate the resource contention in the data preparation stages. Thus, all the invocations could finish the computation in a short time, which contributes to higher system throughput.

It should be noted that \sota{}-F shows worse performance compared with \sota{} both in latency and throughput. This is because \sota{} does not concern the resource contention in the data preparation stages. While \sota{}-F could launch more function invocations on the GPU, it suffers more serious and unrestrained contention of the data-loading paths than \sota{}.

\autoref{fig:ratio} illustrates the memory usage of \sota{}, \sota{}-F, DGSF, and \sysname{} during this period. The average memory usage of \sysname{} only accounts for 18.7\% of \sota{}, 21.7\% of \sota{}-F, and 37.5\% of DGSF. The improved memory usage also comes from the memory sharing of \sysname{}. \sysname{} could achieve better latency and throughput performance using at most 37.5\% memory of the baseline systems. Note that while \sysname{} uses less memory, it could hold more cold functions on the GPU, thus further improving the function latency performance.

\begin{figure}
\centering
\includegraphics[width=\columnwidth]{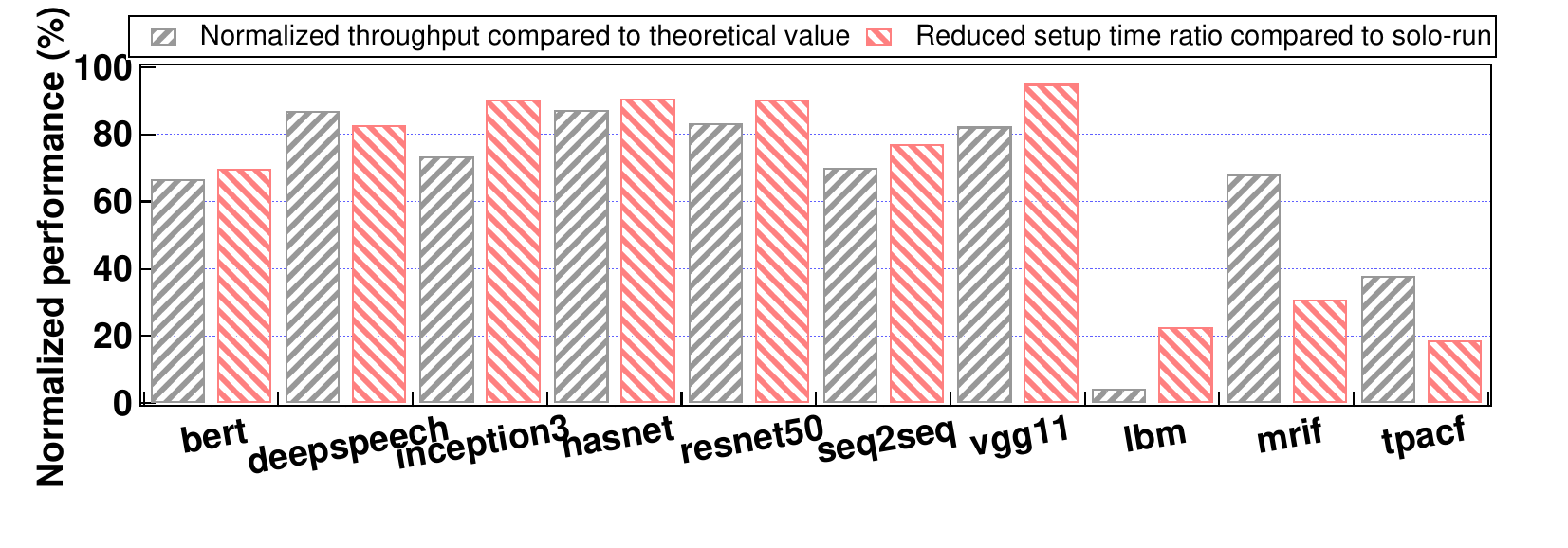}
\caption{The Normalized throughput performance and setup time reduction using \sysname{} compared with \sota{}.}
\label{fig:exp2-theo}
\end{figure}



\subsection{Supported Peak Results}
\label{sec:exp-moti}

\begin{figure*}
	\centering
	\includegraphics[width=2\columnwidth]{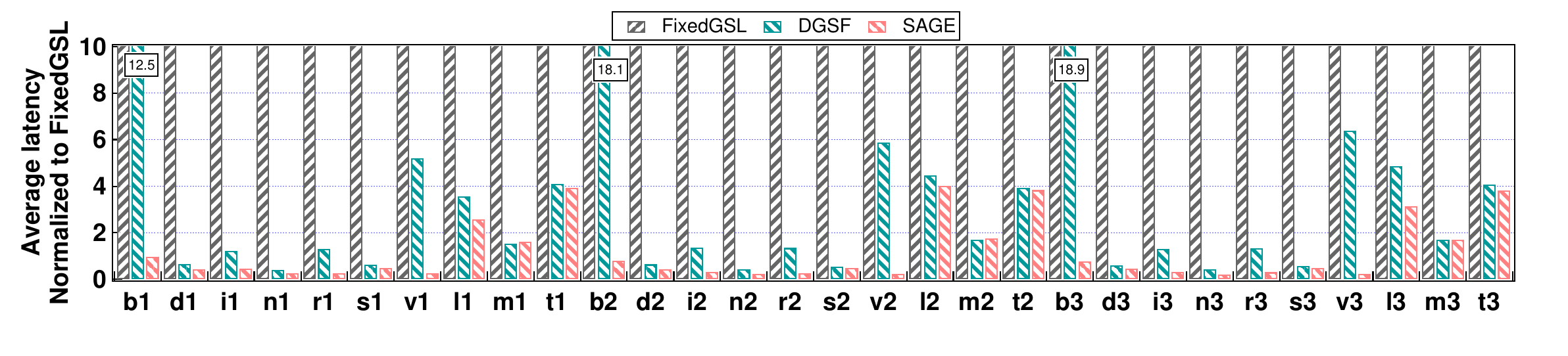}
	\caption{The normalized GPU function's average latency with \sota{}, DGSF, and \sysname{} Compared with \sota{} While 30 functions are serving.  The $x$-axis shows the 30 GPU functions. For instance, $b1$, $b2$, and $b3$ represent the three bert functions. While the \sota{} is the baseline, all the latency results (the gray bar on the far left) with \sota{} is 100.}
	\label{fig:exp3_latency}
\end{figure*}

In this section, we use the same experimental setup as \S~\ref{sec:moti} to evaluate the theoretical throughput ratio for \sysname{}. The gray bars in \autoref{fig:exp2-theo} shows the throughput ratio of each application using \sysname{}. On average, the system throughput is 65.1\% of its theoretical value. These results demonstrate that \sysname{} has better performance than the baseline systems. The improved performance of \sysname{} comes from the fast setup and memory sharing of function invocations. 

The red bars in \autoref{fig:exp2-theo} show the normalized reduced time of each application using \sysname{} compared to the solo-run case. \sysname{} reduces the setup time of function invocations by 66.8\% on average. While the function invocation could get the data ready in a short period, the system throughput could be improved. Note that, there is a significant gap between \sysname{} and the theoretical value because of the resource contention. Although \sysname{} could reduce the data loading pressure by sharing the read-only memory in the GPU function, some data preparation could not be avoided. Furthermore, some tasks like $slbm$ could not benefit from the read-only memory sharing. 

\subsection{Efficiency with More functions}
\label{sec:more-func}

In this subsection, we use 30 functions to further demonstrate the effectiveness of \sysname{}. While the system is required to handle more functions, the memory-sharing capability between multiple invocations of the same function is reduced. Specifically, we use 10 benchmarks in Section 7.1 to simulate 30 benchmarks. For example, we use $bert1$, $bert2$, $bert3$ to represent different benchmarks.

\autoref{fig:exp3_latency} shows the average latency of 30 functions under \sota{}, DGSF, and \sysname{}. On average, \sysname{} outperforms \sota{} and DGSF by 211.9$\times$ and 5.9$\times$, respectively. Additionally, \sysname{} outperforms \sota{} and DGSF with the tail latency by 49.6$\times$ and 4.6$\times$, respectively. While more functions on one GPU node mean a low load for each function, the memory-sharing opportunity between multiple invocations of the same function is reduced. However, \sysname{} still could rely on the parallel setup method and multi-stage resource exit method to optimize the function latency. With these two methods, the function's invocations could still benefit from reduced setup time and a warm context environment.

At the same time, experimental results also show that \sysname{} outperforms \sota{} and DGSF in terms of system throughput by 9.7$\times$ and 1.19$\times$, respectively. This also comes from the possible read-only memory sharing and multi-stage resource exit methods. While the memory sharing opportunity is reduced, \sysname{} could still utilize it to improve the system throughput. In addition, the multi-stage resource exit mechanism could keep the resources warm for a longer time, which also helps to improve the system throughput.

\subsection{Effectiveness of parallelized setup mechnaism}
\label{sec:abalation0}

\begin{figure}
\centering
\includegraphics[width=\columnwidth]{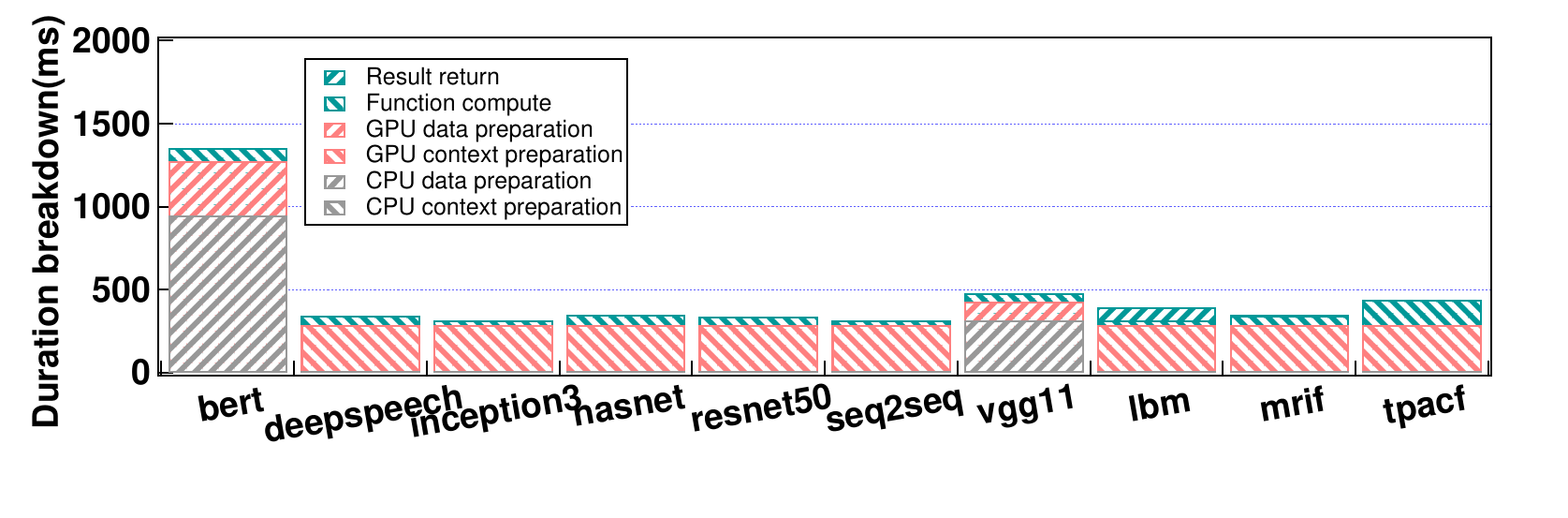}
\caption{The duration breakdown using the parallelized function setup mechanism only.}
\label{fig:exp-8}
\end{figure}

To evaluate the effectiveness of the parallelized function setup mechanism, we measure the end-to-end duration of the GPU functions like the experiment in \S\ref{sec:moti_long}. The function invocations are generated in a close-loop manner.

\autoref{fig:exp-8} shows the duration breakdown of all the benchmark functions with \sysname{}. Compared with the results in \autoref{fig:moti_duration}, \sysname{} reduces the setup time of all the functions by 20.8\% on average. This result could also be observed from the figure. As for the benchmark functions, they either have the GPU context preparation time or the data preparation time. This is because the parallelized function setup mechanism hides another setup stage, which improves the function setup performance. Benefiting from the parallelized function setup mechanism, \sysname{} could improve the 99\%-ile latency of all the benchmark functions.

\subsection{Effectiveness of multi-stage mechnaism}
\label{sec:abalation1}

\begin{table}
\caption{Latency breakdown of $resnet50$ running with multi-stage resource exit mechanism.}
\label{tb:stage}
\centering
\scriptsize
\begin{tabular}{c|c|c|c|c|c|c}
\hline
\textbf{Time (ms)}  & \textbf{Baseline} & \textbf{stage 1} & \textbf{stage 2} & \textbf{stage 3} & \textbf{stage 4} & \textbf{cold}  \\ \hline
end-to-end & 399.4  & 28.9  & 49.7  & 309.5  & 309.5  & 310.5  \\ \hline
return data  & 0.1  & 0.1  & 0.1  & 0.1  & 0.1  & 0.1  \\ \hline
compute  & 24.3  & 24.3  & 24.3  & 24.3  & 24.3  & 24.3  \\ \hline
GPU data  & 21.7  & 0.9  & 21.7  &\textbf{21.7}  & \textbf{21.7}  & \textbf{21.7} \\ \hline
GPU context  & 285.1  & 0  & 0  & \textit{285.1}  & \textit{285.1}  & \textit{285.1}  \\ \hline
CPU data  & 67.2  & 3.6  & 3.6  & \textbf{3.6}  & \textbf{67.2}  & \textbf{67.2} \\ \hline
CPU context  & 1  & 0  & 0  & 0  & 0  & 1  \\ \hline
\end{tabular}
\end{table}

As shown in \autoref{tb:stage}, the invocations could be in four stages or in a cold state. We use Resnet50 as the representative benchmark to show the breakdown in different stages. Compared with the traditional method which only includes warm state and cold state, SAGE contains five stages for fine-grained resource management. As shown in the table, the bold values and the zero values are hidden by the Italic value in the mechanism. For example, the GPU data loading time (67.2) and the CPU data loading time (21.7) are hidden by the GPU context creation time (285.1) in stage 4.

SAGE improves the function’s latency by an average of 6.1$\times$ in the other four states compared to the cold state (baseline), with a minimum of 1.3$\times$. All other benchmarks show similar results. The experimental results show that SAGE reduces 91.7\% of the queries in the cold state. Therefore, the multi-stage resource exit mechanism could improve the function’s 99\%-ile latency.

\subsection{Effectiveness of read-only memory sharing}
\label{sec:abalation2}

In this subsection, we disable the read-only memory sharing of \sysname{} (denoted as \sysname{}-NR), and compare its latency and system throughput with DGSF and \sysname{}. In this experiment, we extend DGSF to destroy the context while there are no function invocations in 30 seconds. 

\begin{figure}
	\centering
	\includegraphics[width=\columnwidth]{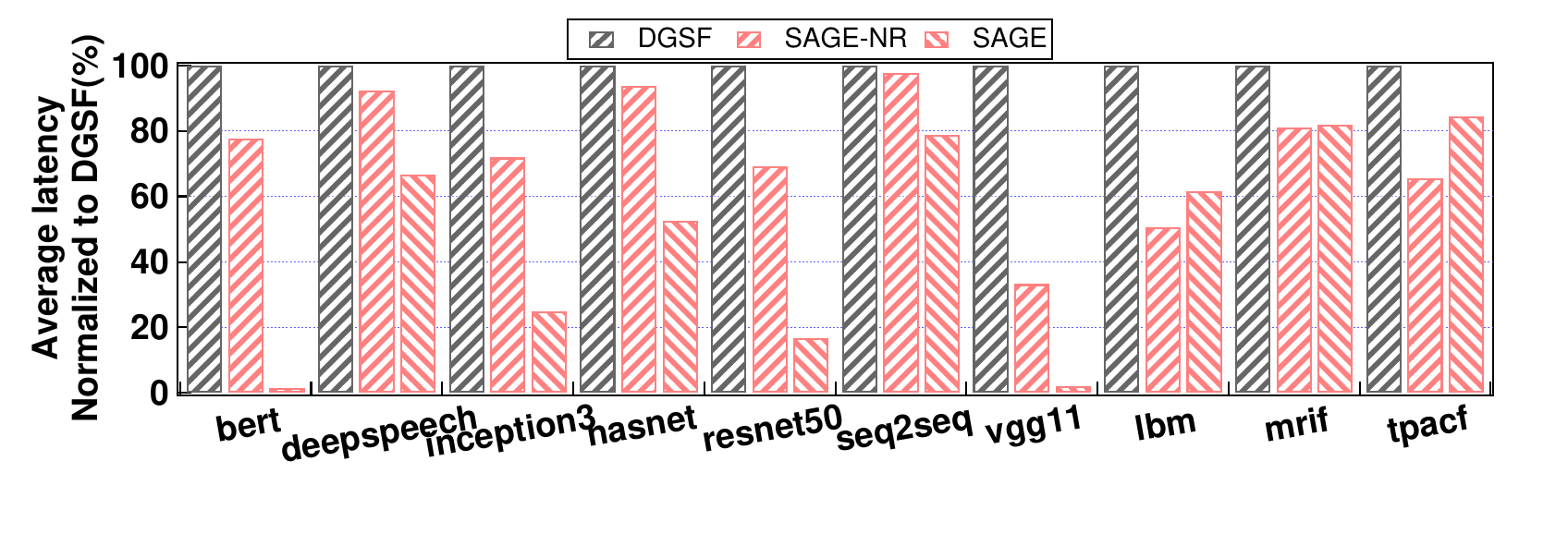}
	\caption{The normalized GPU function's average latency with DGSF, \sysname{}-NR and \sysname{} Compared with DGSF.}
	\label{fig:exp-6}
\end{figure}

\autoref{fig:exp-6} shows the average latency of 10 functions under DGSF, \sysname{}-NR, and \sysname{}. On average, \sysname{} outperforms \sysname{}-NR and DGSF by 8.2$\times$ and 13.3$\times$, respectively. Without the read-only memory sharing method, \sysname{}-NR still needs to suffer resource contention in the data loading paths. Meanwhile, although DGSF could eliminate the context creation time using pre-created context, it still suffers from context creation time when there is no created context. While \sysname{}-NR could utilize the multi-stage resource exit and parallelized setup preparation methods, \sysname{}-NR could perform better than DGSF. Furthermore, \sysname{}-NR has a similar memory usage with DGSF. \sysname{}-NR's memory usage is only 38.3\% of \sysname{}.
s

\subsection{Scaling \sysname{} Out}
\label{sec:scale_out}

\begin{figure}
	\centering
	\includegraphics[width=\columnwidth]{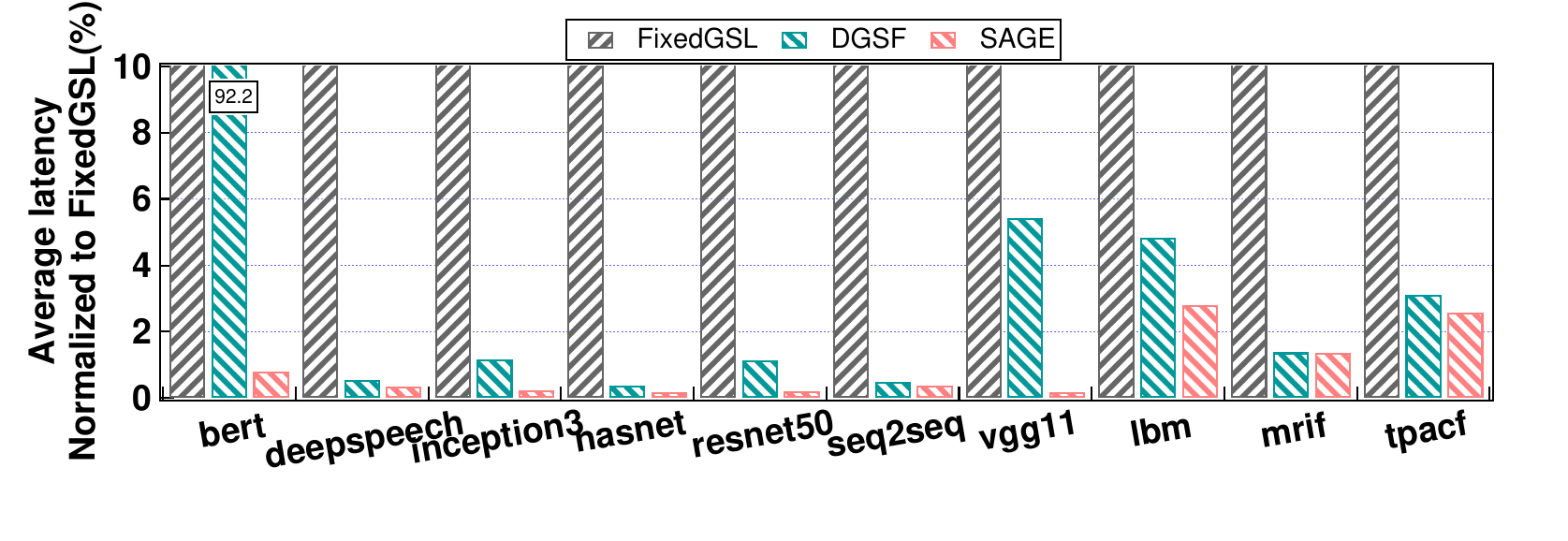}
	\caption{The GPU function's average latency with \sota{}, DGSF and \sysname{} Compared with \sota{} under the cluster with 4 A100. While the \sota{} is the baseline, all the latency results (the gray bar on the far left) with \sota{} is 100.}
	\label{fig:exp-scale}
\end{figure}

To assess \sysname{}'s scalability, we conduct the performance tests on a cluster with 4 A100 devices on 4 nodes. \autoref{fig:exp-scale} illustrates the average latency of ten applications using \sota{}, DGSF, and \sysname{} on the cluster. On average, \sysname{} surpasses \sota{} and DGSF by 207.1$\times$ and 12.5$\times$, respectively. Experimental results also show that \sysname{} delivers a 10.3$\times$ and 1.18$\times$ increase in system throughput compared to \sota{}, and DGSF, respectively.

Based on the aforementioned data, it can be deduced that \sysname{} has excellent scalability. Although we dispatch the function invocations to GPU randomly, \sysname{} could still improve the overall system's average latency and system throughput. This is because the proposed optimization methods of \sysname{} are orthogonal to the task distribution strategy at the cluster level. \sysname{} can integrate any cluster-level scheduling strategies proposed for specific scenarios to further achieve performance improvements.

\subsection{Overhead of \sysname{}}
\label{sec:overhead}

The overhead of \sysname{} arises from the runtime interception of functions and the communication between the kernel executor and memory daemon. Since the runtime interception of functions has been widely used in the past decade, its overhead is very small. At the same time, we use shared memory to support the communication between the executor and daemon. The communication overhead is microsecond-level. Therefore, the overhead of \sysname{} is negligible.

\subsection{Other Lessons Learned}
We have got some other lessons during the research.

{\bf Lesson-1: The fine-grained GPU memory management also enables the fine-grained pricing model for GPU serverless.}
While memory is crucial for GPUs, it is more reasonable to charge the users based on both the memory usage (the size of memory $\times$ hours) and computation usage. In this case, the pricing model of traditional \sota{} is not reasonable as a GPU function may not need 1GB memory. 

{\bf Lesson-2: While a GPU often has many streaming multiprocessors (SMs), \sysname{} can be extended to explicitly allocate SM allocation to GPU functions or adjust the processing order of GPU serverless functions.} 
The functionality can be done through the shim module and provides better control over the progress of concurrent GPU functions.

{\bf Lesson-3: While SAGE's fine-grained memory management enables GPU to have more memory available, it could be extended to use a cache replacement strategy for container warm-up.}  For a new-coming function, we can use a cache replacement strategy to replace long-inactive GPU functions. Meanwhile, while there are no new GPU functions arriving, existing GPU functions could always reside in GPU memory.

%% file: contents/conclusion.tex
\section{Conclusion\label{sec:conclusion}}

In this paper, we propose \sysname{}, an efficient serverless system with fast setup and high throughput for GPU functions. \sysname{} proposes two innovative mechanisms to enhance the function latency and throughput performance for the GPU serverless system, which are parallelized function setup and sharing-based memory management. The parallelized function setup mechanism enables the fast function setup by parallelizing data preparation and context creation. Meanwhile, the sharing-based memory management allows multiple invocations of the same function to share the read-only memory and context memory. This reduces the resource contention in the datapaths and then improves the system throughput. Our experimental results show that \sysname{} improves function throughput by $1.22 \times$ and reduces function duration by $11.3 \times$, compared with state-of-the-art solutions.